\documentclass{sig-alternate}
\usepackage{graphicx}\usepackage[caption=false]{subfig}
\usepackage{epstopdf}
\usepackage[table]{xcolor}
\usepackage{multirow}
\usepackage{color}
\usepackage{url}   
\usepackage{balance}
\usepackage{fancybox}
\usepackage{booktabs} 
\usepackage{algorithm}\usepackage{algorithmic}
\usepackage{amsmath}\usepackage{amsfonts,amssymb}
\usepackage{bm}
\usepackage{comment}
\usepackage{cases}
\usepackage{threeparttable} 
\usepackage[bookmarks=false]{hyperref}
\hypersetup{
	hidelinks=true,
}

\begin{document}
	\setcopyright{acmcopyright}
	\conferenceinfo{FSE'16}{Nov 13--18, 2016, Seattle, WA, USA}
	\title{Deep API Learning}	
	
	

	\numberofauthors{1}
	\author{
		\alignauthor
		Xiaodong Gu$^1$, Hongyu Zhang$^2$, Dongmei Zhang$^3$, and Sunghun Kim$^1$\\
		\affaddr{$^1$The Hong Kong University of Science and Technology, Hong Kong, China}\\
		\email{guxiaodong1987@126.com hunkim@cse.ust.hk}\\
		\affaddr{$^2$The University of Newcastle, Callaghan, Australia}\\
		\email{hongyu.zhang@newcastle.edu.au}\\
		\affaddr{$^3$Microsoft Research, Beijing, China}\\
		\email{dongmeiz@microsoft.com}
	}
	
	\maketitle
	
	\begin{abstract}
		
		Developers often wonder how to implement a certain functionality (e.g., how to parse XML files) using APIs. Obtaining an API usage sequence based on an API-related natural language query is very helpful in this regard. Given a query, existing approaches utilize information retrieval models to search for matching API sequences. These approaches treat queries and APIs as bags-of-words and lack a deep understanding of the semantics of the query.
		
		We propose \textsc{DeepAPI}, a deep learning based approach to generate API usage sequences for a given natural language query. Instead of a bag-of-words assumption, it learns the sequence of words in a query and the sequence of associated APIs.
		\textsc{DeepAPI} adapts a neural language model named RNN Encoder-Decoder. It encodes a word sequence (user query) into a fixed-length context vector, and generates an API sequence based on the context vector. We also augment the RNN Encoder-Decoder by considering the importance of individual APIs.
		We empirically evaluate our approach with more than 7 million annotated code snippets collected from GitHub. The results show that our approach generates largely accurate API sequences and outperforms the related approaches.
	\end{abstract}

	%
	
	\section{Introduction}	
	
	To implement a certain functionality, for example, how to parse XML files, developers often reuse existing class libraries or frameworks by invoking the corresponding APIs. Obtaining which APIs to use, and their usage sequence (the method invocation sequence among the APIs) is very helpful in this regard~\cite{fowkes2015parameter,upminer,mapo2006}. 
	For example, to ``\textit{parse XML files}'' using JDK library, the desired API usage sequence is as follows: 
	
	~~~~~$DocumentBuilderFactory.newInstance$ 
	
	~~~~~$DocumentBuilderFactory.newDocumentBuilder$  
	
	~~~~~$DocumentBuilder.parse$ 
	
	Yet learning the APIs of an unfamiliar library or software framework can be a significant obstacle for developers~\cite{fowkes2015parameter,upminer}. A large-scale software library such as .NET framework and JDK could contain hundreds or even thousands of APIs. In practice, usage patterns of API methods are often not well documented~\cite{upminer}. 
	In a survey conducted by Microsoft in 2009, 67.6\% respondents mentioned that there are obstacles caused by inadequate or absent resources for learning APIs~\cite{Robillard09}. Another field study found that a major challenge for API users is to discover the subset of the APIs that can help complete a task~\cite{Robillard2010}. 
	
	A common place to discover APIs and their usage sequence is from a search engine. Many developers search APIs from general web search engines such as Google and Bing. Developers can also perform a code search over an open source repository such as GitHub~\cite{githubsearch} and then utilize an API usage pattern miner~\cite{fowkes2015parameter,upminer,mapo2006} to obtain the appropriate API sequences.
	
	However, search engines are often inefficient and inaccurate for programming tasks~\cite{Stylos06}. General web search engines are not designed to specifically support programming tasks. Developers need to manually examine many web pages to learn about the APIs and their usage sequence. Besides, most of search engines are based on keyword matching without considering the semantics of natural language queries~\cite{holmes2009end}. It is often difficult to discover relevant code snippets and associated APIs. 
	
	Recently, Raghothaman et al.~\cite{swim} proposed SWIM, which translates a natural language query to a list of possible APIs using a statistical word alignment model~\cite{smt}. SWIM then uses the API list to retrieve relevant API sequences. 
	However, the statistical word alignment model it utilizes is based on a bag-of-words assumption without considering the sequence of words and APIs. Therefore, it cannot recognize the deep semantics of a natural language query. For example, as described in their paper~\cite{swim}, it is difficult to distinguish the query \emph{convert int to string} from \emph{convert string to int}. 

	To address these issues, we propose DeepAPI, a novel, deep-learning based method that generates relevant API usage sequences given a natural language query. We formulate the API learning problem as a machine translation problem: given a natural language query $x=x_1,...,x_N$ where $x_i$ is a keyword, we aim to translate it into an API sequence $y=y_1,...,y_T$ where $y_j$ is an API. 	
	\textsc{DeepAPI} shows a deep understanding of natural language queries in two aspects: 
	\begin{itemize}
		\vspace{-0.2\baselineskip}
		\item	First, instead of matching keywords, \textsc{DeepAPI} learns the semantics of words by embedding them into a vector representation of context, so that semantically related words can be recognized.
		\vspace{-0.2\baselineskip} 
		\item	Second, instead of word-to-word alignment, \textsc{DeepAPI} learns the sequence of words in a natural language query and the sequence of associated APIs. It can distinguish the semantic differences between queries with different word sequences.
		\vspace{-0.2\baselineskip}
	\end{itemize}	
	
	\textsc{DeepAPI} adapts a neural language model named RNN Encoder-Decoder~\cite{cho2014phrase}. Given a corpus of annotated API sequences, i.e., \textsf{$\langle$API sequence, annotation$\rangle$} pairs, \textsc{DeepAPI} trains the language model that encodes each sequence of words (annotation) into a fixed-length context vector and decodes an API sequence based on the context vector. 
	Then, in response to an API-related user query, it generates API sequences by consulting the neural language model. 
	
	To evaluate the effectiveness of \textsc{DeepAPI}, we collect a corpus of 7 million annotated code snippets from GitHub. We select 10 thousand instances for testing and the rest for training the model. After 240 hours of training (1 million iterations), we measure the accuracy of \textsc{DeepAPI} using BLEU score~\cite{bleu}, a widely used accuracy measure for machine translation.
	Our results show that \textsc{DeepAPI} achieves an average BLEU score of 54.42, outperforming two related approaches, that is, code search with pattern mining (11.97) and SWIM~\cite{swim} (19.90). 
	We also ask \textsc{DeepAPI} 30 API-related queries collected from real query logs and related work. On average, the rank of the first relevant result is 1.6. 80\% of the top 5 returned results and 78\% of the top 10 returned results are deemed relevant. Our evaluation results confirm the effectiveness of \textsc{DeepAPI}.
	
	
	The main contributions of our work are as follows:
	\begin{itemize}
		\vspace{-0.2\baselineskip}
		\item To our knowledge, we are the first to adapt a deep learning technique to API learning. Our approach leads to more accurate API usage sequences as compared to the state-of-the-art techniques.
		\vspace{-0.2\baselineskip}
		\item 
		We develop \textsc{DeepAPI}\footnote{available at: \url{https://guxd.github.io/deepapi/}}, a tool that generates API usage sequences based on natural language queries. We empirically evaluate \textsc{DeepAPI}'s accuracy using a corpus of 7 million annotated Java code snippets.  
	\end{itemize}
	
	The rest of this paper is organized as follows. Section~\ref{s:background} describes the background of the deep learning based neural language model. 
	Section~\ref{s:tech} describes the application of the RNN Encoder-Decoder, a deep learning based neural language model, to API learning. Section~\ref{s:approach} describes the detailed design of our approach. Section~\ref{s:eval} presents the evaluation results. Section~\ref{s:discuss} discusses our work, followed by Section~\ref{s:related} that presents the related work. We conclude the paper in Section~\ref{s:conclusion}.

	%
	%
	
	\section{Deep Learning for Sequence Generation}\label{s:background}
	

	Our work adopts and augments recent advanced techniques from deep learning and neural machine translation~\cite{bahdanau2014attention,cho2014phrase,sutskever2014seq2seq}. These techniques are on the basis of Sequence-to-Sequence Learning~\cite{sutskever2014seq2seq}, namely, generating a sequence (usually a natural language sentence) conditioned on another sequence. In this section, we discuss the background of these techniques.
		
	\subsection{Language Model}
	It has been observed that software has naturalness~\cite{naturalness}. Statistical language models have been adapted to many software engineering tasks~\cite{allamanis2014generative} such as learning natural code conventions~\cite{naturalize}, code suggestion~\cite{localness}, and code completion~\cite{rnncodecompletion}. These techniques regard source code as a special language and analyze it using statistical NLP techniques. 
	
	The language model is a probabilistic model of how to generate sentences in a language. 
	It tells how likely a sentence would occur in a language.
	For a sentence~$y$, where $y=(y_1,...,y_T)$ is a sequence of words, the language model aims to estimate the joint probability of its words~$Pr(y_1,...,y_T)$.\\
	Since 
	\vspace{-0.4\baselineskip}
	\begin{equation}\label{eq:lm}
	\small
	\vspace{-0.4\baselineskip}
	Pr(y_1,...,y_T)= \prod_{t=1}^T Pr(y_t|y_1,...,y_{t-1})
	\end{equation}
	it is equivalent to estimate the probability of each word in $y$ given its previous words, namely, what a word might be given its predecessing words. 
	
	As $Pr(y_t|y_1,...,y_{t-1})$ is difficult to estimate, most applications use ``n-gram models''~\cite{ngram} to approximate it, that is,
	\begin{equation}
	Pr(y_t|y_1,...,y_{t-1})\backsimeq Pr(y_t|y_{t-n+1},...,y_{t-1})
	\end{equation}
	where an n-gram is defined as $n$ consecutive words. This approximation means that the next word~$y_t$ is conditioned only on the previous $n-1$ words.
	
	\subsection{Neural Language Model}
\begin{figure}[tb]
			\setlength{\abovecaptionskip}{0pt}
			\setlength{\belowcaptionskip}{0pt}
			\centering
			\subfloat[\scriptsize RNN Structure]{
				\includegraphics[width=1in]{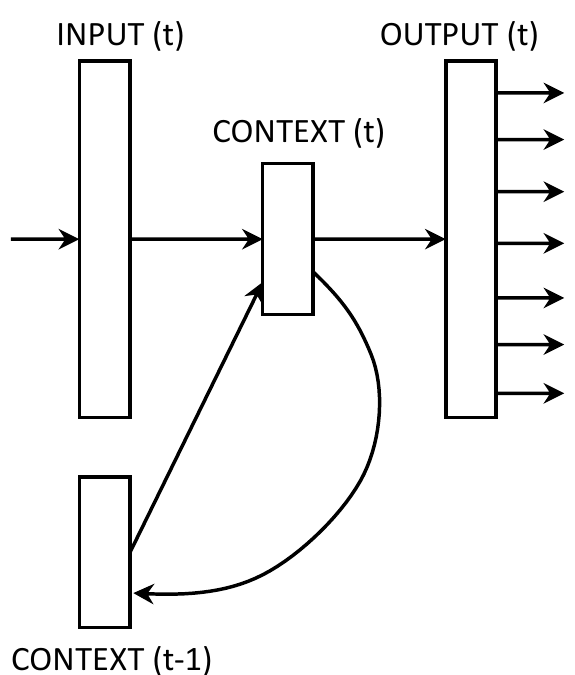}
				\label{fig:approach:rnn}} 
			\space{  } 
			\subfloat[\scriptsize RNNLM for sentence estimation]{
				\includegraphics[width=2.05in]{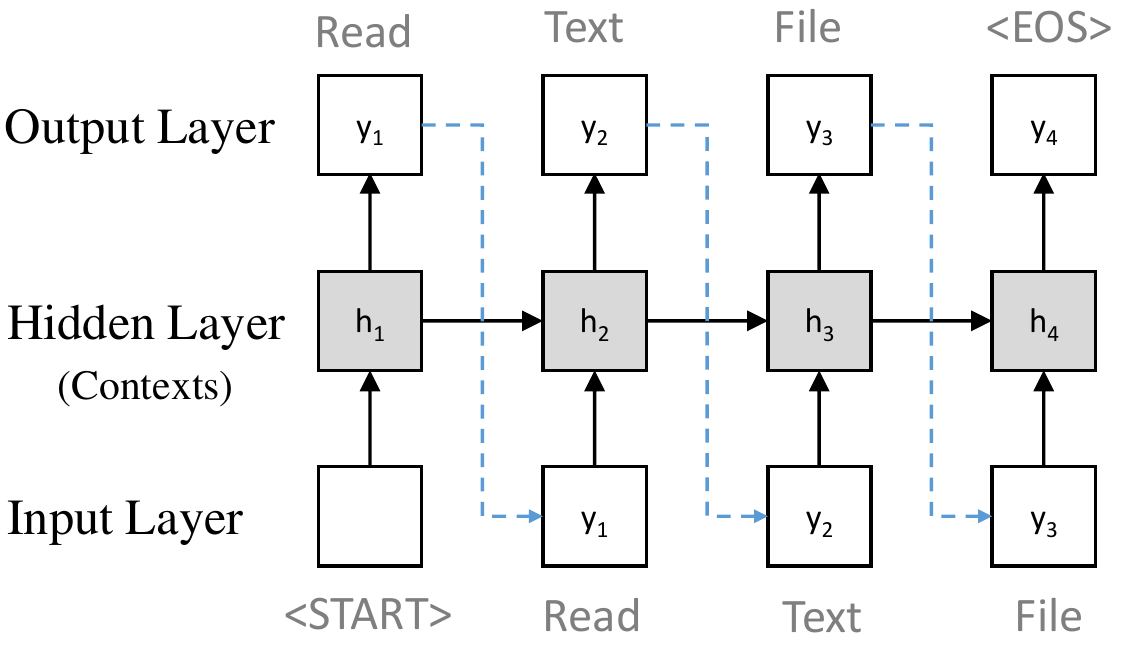}
				\label{fig:approach:rnnlm}}
			\vspace{-8pt} 
			\caption{\small Illustration of the RNN Language Model}
			\label{fig:approach:rnnillus}
			\vspace{-1\baselineskip}
		\end{figure}

	The neural language model is a language model based on neural networks. Unlike the n-gram model which predicts a word based on a fixed number of predecessing words, a neural language model can predict a word by predecessing words with longer distances. It is also powerful to learn distributed representations of words, i.e, word vectors~\cite{wordvec}. 
	We adopt RNNLM~\cite{mikolov2010rnnlm}, a language model based on a deep neural network, that is, Recurrent Neural Network~(RNN)~\cite{mikolov2010rnnlm}. 
	Figure~\ref{fig:approach:rnn} shows the basic structure of an RNN. The neural network includes three layers, that is, an input layer which maps each word to a vector, a recurrent hidden layer which recurrently computes and updates a hidden state after reading each word, and an output layer which estimates the probabilities of the following word given the current hidden state. 
	
	Figure~\ref{fig:approach:rnnlm} shows an example of how RNNLM estimates the probability of a sentence, that is, the probability of each word given predecessing words (Equation~\ref{eq:lm}). To facilitate understanding, we expand the recurrent hidden layer for each individual time step.
	The RNNLM reads the words in the sentence one by one, and predicts the possible following word at each time step.
	At step~$t$, it estimates the probability of the following word~$p(y_{t+1}|y_1,...,y_t)$ by three steps:
	First, the current word~$y_t$ is mapped to a vector~$\bm{y}_t$ by the input layer:
	\begin{equation}
		\small
		\bm{y}_t=input(y_t)
	\end{equation}
	Then, it generates the hidden state (values in the hidden layer)~$\bm{h}_t$ at time~$t$ according to the previous hidden state $\bm{h}_{t-1}$ and the current input $\bm{y}_t$:
	\begin{equation}
		\small
		\bm{h}_t=f(\bm{h}_{t-1},\bm{y}_t)
	\end{equation}
	Finally, the $Pr(y_{t+1}|y_1,...,y_t)$ is predicted according to the current hidden state~$\bm{h}_t$:
	\begin{equation}
		\small
		Pr(y_{t+1}|y_1,...,y_t)=g(\bm{h}_t) 
		\vspace{-0.2\baselineskip}
	\end{equation}
	
	During training, the network parameters are learned from data to minimize the error rate of the estimated $y$ (details are in~\cite{mikolov2010rnnlm}).
	

	\subsection{RNN Encoder-Decoder Model}
	The RNN Encoder-Decoder~\cite{cho2014phrase} is an extension of the basic neural language model (RNNLM). It assumes that there are two languages, a source language and a target language. It generates a sentence~$y$ of the target language given a sentence~$x$ of the source language. To do so, it first summarizes the sequence of source words~$x_1,...,x_{T_x}$ into a fixed-length context vector: 
	\vspace{-0.1\baselineskip}
	\begin{equation}\label{eq:rnnencdec:h}
		\small
		\bm{h}_t=f(\bm{h}_{t-1},\bm{x}_t)
		\vspace{-0.2\baselineskip}
	\end{equation}
	and
	\begin{equation}\label{eq:rnnencdec:c}
		\small
		\vspace{-0.2\baselineskip}
		\bm{c}=\bm{h}_{T_x}
	\end{equation}
	where $f$ is a non-linear function that maps a word of source language~$\bm{x}_t$ into a hidden state~$\bm{h}_t$ at time~$t$ by considering the previous hidden state~$\bm{h}_{t-1}$. The last hidden state~$\bm{h}_{T_x}$ is selected as a context vector~$\bm{c}$.
	
	Then, it generates the target sentence~$y$ by sequentially predicting a word~$y_t$ conditioned on the source context~$\bm{c}$ as well as previous words $y_1,...,y_{t-1}$:
	\vspace{-0.4\baselineskip}
	\begin{equation}\label{eq:encdec:pred}
		\small
		\vspace{-0.4\baselineskip}
		Pr(y)=\prod_{t=1}^T p(y_t|{y_1,...,y_{t-1}},\bm{c})
	\end{equation} 
	
	The above procedures, i.e., $f$ and $p$ can be represented using two recurrent neural networks respectively, an encoder RNN which learns to transform a variable length source sequence into a fixed-length context vector, and a decoder RNN which learns a target language model and generates a sequence conditioned on the context vector. 
	The encoder RNN reads the source words one by one. At each time stamp~$t$, it reads one word, then updates and records a hidden state. When reading a word, it computes the current hidden state~$\bm{h}_t$ using the current word~$\bm{x}_t$ and the previous hidden state~$\bm{h}_{t-1}$. When it finishes reading the end-of-sequence word \emph{<EOS>}, it selects the last hidden state~$\bm{h}_{T_x}$ as a context vector~$\bm{c}$. 
	The decoder RNN then sequentially generates the target words by consulting the context vector (Equation~\ref{eq:encdec:pred}). 
	It first sets the context vector as an initial hidden state of the decoder RNN. At each time stamp~$t$, it generates one word based on the current hidden state and the context vector. Then, it updates the hidden state using the generated word (Equation~\ref{eq:rnnencdec:h}). It stops when generating the end-of-sentence word \emph{<EOS>}.
	
	The RNN Encoder-Decoder model can then be trained to maximize the conditional log-likelihood~\cite{cho2014phrase}, namely, minimize the following objective function:
	\vspace{-0.4\baselineskip}
	\begin{equation}\label{eq:rnnenecdec:obj}
	   \small
		\mathcal{L}(\theta)=\frac{1}{N}\sum_{i=1}^N\sum_{t=1}^{T}\mathrm{cost}_{it}
		\vspace{-0.4\baselineskip}
	\end{equation}
	where $N$ is the total number of training instances, while $T$ is the length of each target sequence. 
	$cost_{it}$ is the cost function for the $t$-th target word in instance~$i$. It is defined as the negative log likelihood: 
	\vspace{-0.2\baselineskip}
	\begin{equation}\label{eq:rnnencdec:cost}
		\vspace{-0.2\baselineskip}
		\mathrm{cost}_{it}=-\log p_\theta(y_{it}|x_i)
	\end{equation}
	where $\theta$ denotes model parameters such as weights in the neural network, while $p_{\theta}(y_{it}|x_i)$ (derived from Equation~\ref{eq:rnnencdec:h} to \ref{eq:encdec:pred}) denotes the likelihood of generating the $t$-th target word given the source sequence~$x_i$ in instance~$i$ according to the model parameters~$\theta$.
	Through optimizing the objective function using optimization algorithms such as gradient descendant, the optimum $\theta$ value can be estimated.


	\section{RNN Encoder-Decoder Model For API Learning}\label{s:tech}	
	
	\subsection{Application of RNN Encoder-Decoder to API Learning}\label{ss:tech:attension}
		\begin{figure*} [!tb]
			\setlength{\abovecaptionskip}{0pt}
			\setlength{\belowcaptionskip}{0pt}
			\centering 
			\includegraphics[width=5in]{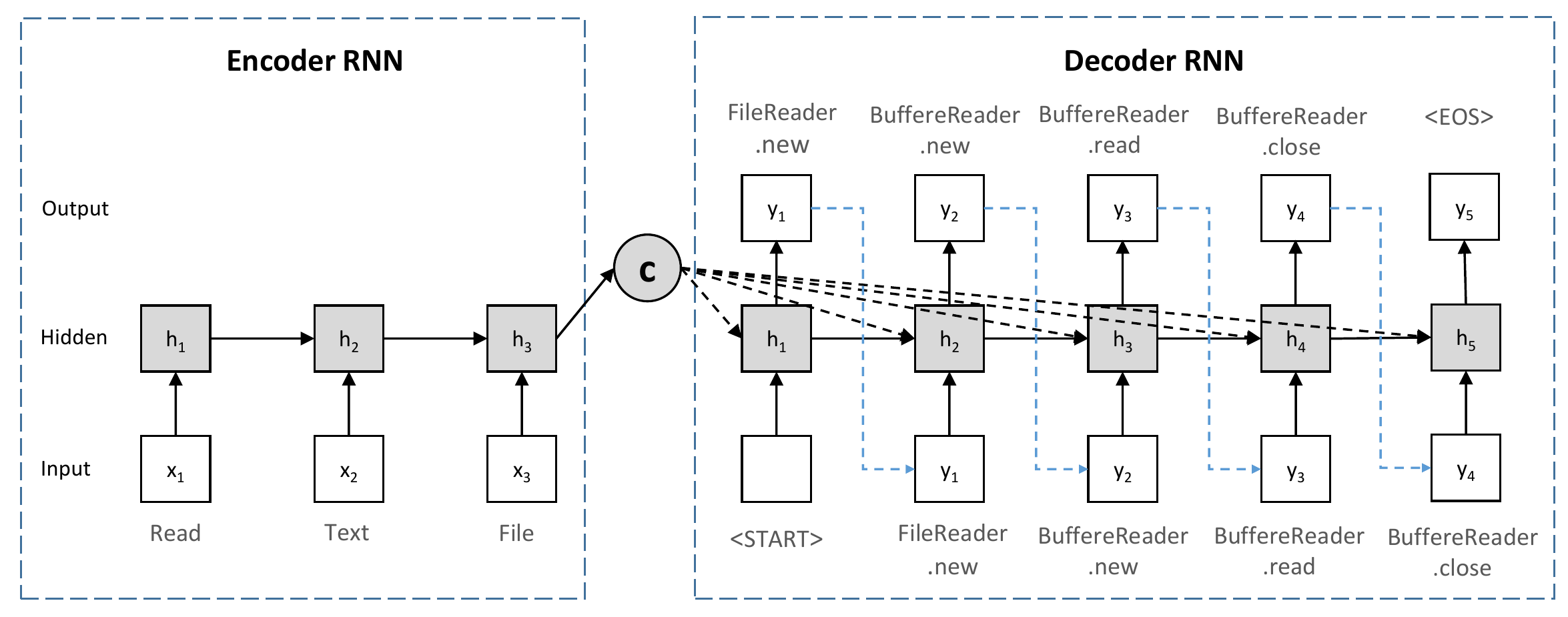} 
			\vspace{-12pt} 
			\caption{\small An Illustration of the RNN Encoder-Decoder Model for API learning}
			\label{fig:approach:rnnencdec}
			\vspace{-1\baselineskip}
		\end{figure*}
	Now we present the idea of applying the RNN Encoder-Decoder model to API learning. We regard user queries as the source language and API sequences as the target language. 
	Figure~\ref{fig:approach:rnnencdec} shows an example of the RNN Encoder-Decoder model for translating a sequence of English words \emph{read text file} to a sequence of APIs. The encoder RNN reads the source words one by one. When it reads the first word~\emph{read}, it embeds the word into vector~$\bm{x}_1$ and computes the current hidden state~$\bm{h}_1$ using $\bm{x}_1$. Then, it reads the second word~\emph{text}, embeds it into $\bm{x}_2$, and updates the hidden state~$\bm{h}_1$ to $\bm{h}_2$ using $\bm{x}_2$. The procedure continues until the encoder reads the last word~\emph{file} and gets the final state~$\bm{h}_3$. The final state~$\bm{h}_3$ is selected as a context vector~$\bm{c}$.
	
	The decoder RNN tries to generate APIs sequentially using the context vector~$\bm{c}$. It first generates \emph{<START>} as the first word~$\bm{y}_0$. Then, it computes a hidden state~$\bm{h}_1$ based on the context vector~$\bm{c}$ and $\bm{y}_0$, and predicts the first API~\emph{FileReader.new} according to $\bm{h}_1$. 
	It then computes the next hidden state~$\bm{h}_2$ according to the previous word vector~$\bm{y}_1$, the context vector~$\bm{c}$, and predicts the second API~\emph{BufferedReader.new} according to $\bm{h}_2$. This procedure continues until it predicts the end-of-sequence word \emph{<EOS>}.
	
	Different parts of a query could have different importance to an API in the target sequence. For example, considering the query \emph{save file in default encoding} and the target API sequence \emph{File.new FileOutputStream.new FileOutputStream.write FileOutputStream.close}, the word \emph{file} is more important than \emph{default} to the target API \emph{File.new}. 
	In our work, we adopt the attention-based RNN Encoder-Decoder model~\cite{bahdanau2014attention}, which is a recent model that selects the important parts from the input sequence for each target word. Instead of generating target words using the same context vector~$\bm{c}$ ($\bm{c}=\bm{h}_{T_x}$), an attention model defines individual $\bm{c}_j$'s for each target word~$y_j$ as a weighted sum of all historical hidden states~$\bm{h}_1,...,\bm{h}_{T_x}$. That is, 
		\vspace{-0.2\baselineskip}
		\begin{equation}
			\small
			\vspace{-0.4\baselineskip}
			\bm{c}_j=\sum_{t=1}^{T_x}\alpha_{jt}\bm{h}_t
		\end{equation} 
	where each $\alpha_{jt}$ is a weight between the hidden state~$\bm{h}_t$ and the target word $y_j$, while $\alpha$ can be modeled using another neural network and learned during training (see details in \cite{bahdanau2014attention}).

	\subsection{Enhancing RNN Encoder-Decoder Model with API importance}\label{ss:tech:enhance}	
	
	The basic RNN Encoder-Decoder model does not consider the importance of individual words in the target sequence either. In the context of API learning, different APIs have different importance for a programming task~\cite{mcmillan2012detecting}. For example, the API~\emph{Logger.log} is widely used in many code snippets. 
	However, it cannot help understand the key procedures of a programming task. Such ubiquitous APIs would be ``weakened'' during sequence generation. 
	
	We augment the RNN Encoder-Decoder model to predict API sequences by considering the individual importance of APIs.
	We define IDF-based weighting to measure API importance as follows:
	\vspace{-0.4\baselineskip}
	\begin{equation}
		w_{idf}(y_t)=log(\frac{N}{n_{y_t}})
		\vspace{-0.4\baselineskip}
	\end{equation}
	where $N$ is the total number of API sequences in the training set and $n_{y_t}$ denotes the number of sequences where the API~$y_t$ appears in the training set. Using IDF, the APIs that occur ubiquitously have the lower weights while the less common APIs have the higher weights.
	
	We use API weight as a penalty term to the cost function (Equation~\ref{eq:rnnencdec:cost}). The new cost function of the RNN Encoder-Decoder model is:
	\vspace{-0.1\baselineskip}
	\begin{equation}\label{eq:newcost}
		\mathrm{cost}_{it}=-\log p_\theta(y_{it}|x_i) - \lambda w_{idf}(y_t)
		\vspace{-0.1\baselineskip}
	\end{equation}
	where $\lambda$ denotes the penalty of IDF weight and is set empirically. 

	\section{DeepAPI: Deep Learning for API sequence generation}\label{s:approach}
\begin{figure} [!tb]
		\setlength{\abovecaptionskip}{0pt}
		\setlength{\belowcaptionskip}{0pt}
		\centering 
		\includegraphics[width=3.4in]{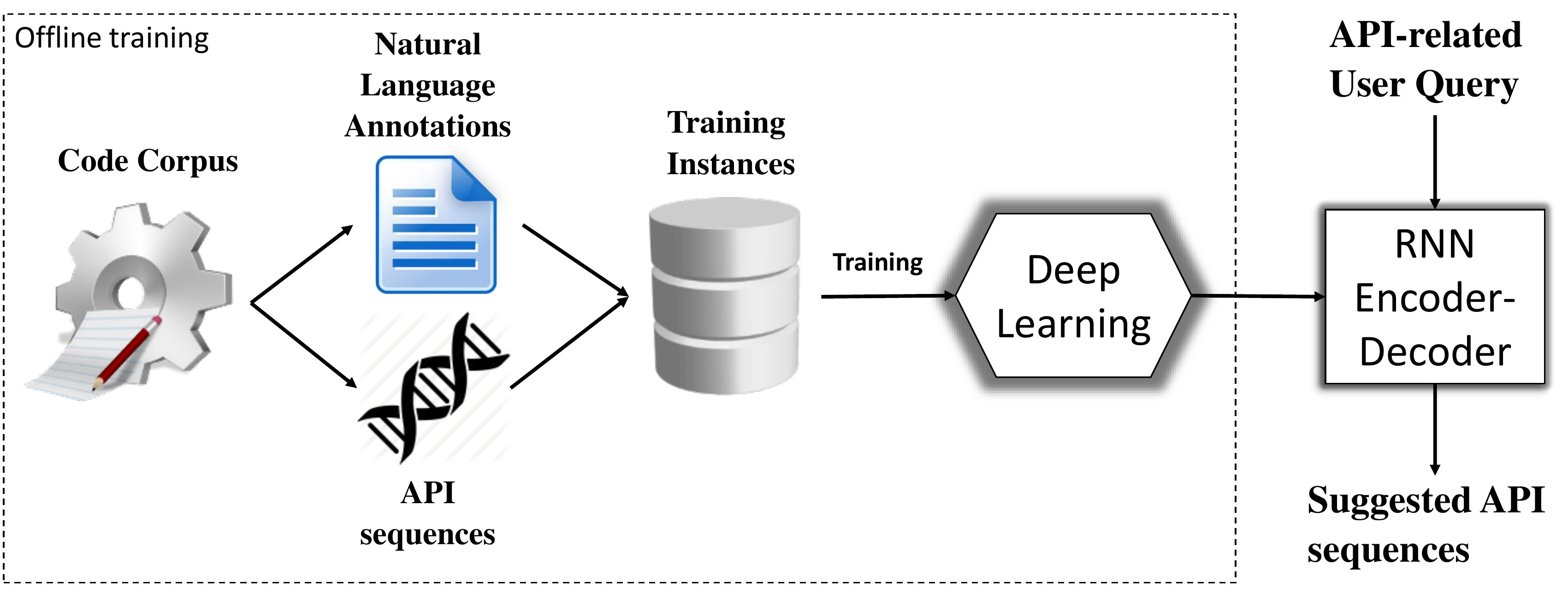} 
		\vspace{-20pt}
		\caption{\small The Overall Workflow of \textsc{DeepAPI}}
		\label{fig:approach:framework}
		\vspace{-1.5\baselineskip}
	\end{figure}

	In this section, we describe \textsc{DeepAPI}, a deep-learning based method that generates relevant API usage sequences given an API-related natural language query. \textsc{DeepAPI} adapts the RNN Encoder-Decoder model for the task of API learning. Figure~\ref{fig:approach:framework} shows the overall architecture of \textsc{DeepAPI}. It includes an offline training stage and an online translation stage. In the training stage, we prepare a large-scale corpus of annotated API sequences (API sequences with corresponding natural language annotations). The annotated API sequences are used to train a deep learning model, i.e., the RNN Encoder-Decoder language model as described in Section~\ref{s:tech}. 
	Given an API-related user query, a ranked list of API sequences can be generated by the language model.
	
	In theory our approach could generate APIs written in any programming languages. In this paper we limit our scope to the JDK library. The details of our method are explained in the following sections.				
	
	\subsection{Gathering a Large-scale API Sequence to Annotation Corpus}\label{ss:approach:data}
	We first construct a large-scale database that contains pairs of API sequences and natural language annotations for training the RNN Encoder-Decoder model.
	We download Java projects from GitHub~\cite{github} created from 2008 to 2014. To remove toy or experimental programs, we only select the projects with at least one star. In total, we collected 442,928 Java projects from GitHub. We use the last snapshot of each project. 
	Having collected the code corpus, we extract \textsf{$\langle$API sequence, annotation$\rangle$} pairs as follows: 
	
	\subsubsection{Extracting API Usage Sequences}
	To extract API usage sequences from the code corpus, we parse source code files into ASTs (Abstract Syntax Trees) using Eclipse's JDT compiler~\cite{jdt}.
	The extraction algorithm starts from the dependency analysis of a whole project repository. We analyze all classes, recording field declarations together with their type bindings. 
	We replace all object types with their real class types. 
	Then, we extract API sequence from individual methods by traversing the AST of the method body:
	\begin{itemize}
		\vspace{-0.5\baselineskip} 
		\item For each constructor invocation $new~C()$, we append the API \textit{C.new} to the API sequence.
		\vspace{-0.5\baselineskip} 
		\item For each method call $o.m()$ where $o$ is an instance of a JDK class $C$, we append the API \textit{C.m} to the API sequence. 
		\vspace{-0.5\baselineskip} 
		\item For a method call passed as a parameter, we append the method before the calling method. For example, $o_1$.$m_1$($o_2$.$m_2$(),$o_3$.$m_3()$), we produce a sequence~$C_2$.$m_2$-$C_3$.$m_3$-$C_1$.$m_1$, where $C_i$ is the JDK class of instance $o_i$.
		\vspace{-0.5\baselineskip} 		
		\item For a sequence of statements $stmt_1$; $stmt_2$ ;...; $stmt_t$, we extract the API sequence $s_i$ from each statement $stmt_i$, concatenate them, and produce the API sequence~$s_1$-$s_2$-...-$s_t$.
		\vspace{-0.5\baselineskip} 
		\item For conditional statements such as if ($stmt_1$) \{ $stmt_2$; \} else \{ $stmt_3$; \}, we create a sequence from all possible branches, that is, $s_1$-$s_2$-$s_3$, where $s_i$ is the API sequence extracted from the statement $stmt_i$. 		
		\vspace{-0.5\baselineskip} 
		\item For loop statements such as while($stmt_1$)\{$stmt_2$;\}, we produce a sequence $s_1$-$s_2$, where $s_1$ and $s_2$ are API sequences extracted from the statement $stmt_1$ and $stmt_2$, respectively. 			
	\end{itemize}
	
	
	\subsubsection{Extracting Annotations}
\begin{figure} [!tb]
		\vspace{4pt}
		\setlength{\belowcaptionskip}{0pt}
		\centering 
		\includegraphics[width=3.2in]{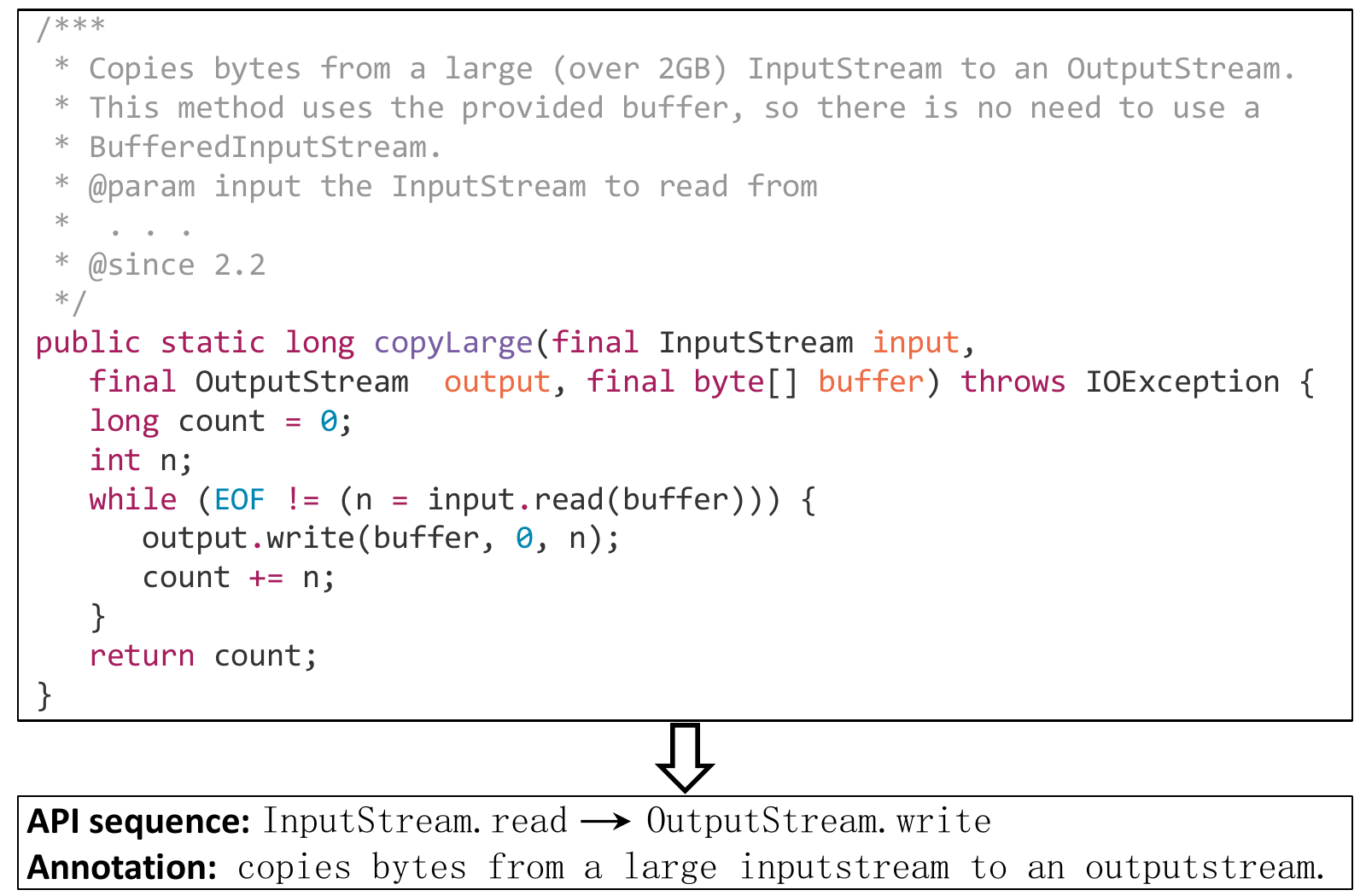} 
		\vspace{-8pt}
		\caption{\small An example of extracting API sequence and its annotation from a Java method $IOUtils.copyLarge$$^{7}$}
		\label{fig:approach:annotation}
		\vspace{-1\baselineskip}
	\end{figure}

	To annotate the obtained API sequences with natural language descriptions, we extract method-level code summaries, specifically, the first sentence of a documentation comment\footnote{A documentation comment in JAVA starts with slash-asterisk-asterisk (/**) and ends with asterisk-slash (*/)} for a method. According to the Javadoc guidance\footnote{\url{http://www.oracle.com/technetwork/articles/java/index-137868.html}}, the first sentence is used as a short summary of a method. Figure~\ref{fig:approach:annotation} shows an example of documentation comment for a Java method $IOUtils.copyLarge$\footnote{\url{https://github.com/apache/commons-io/blob/trunk/src/main/java/org/apache/commons/io/IOUtils.java}} in the Apache commons-io library. 
	
	We use the Eclipse JDT compiler for the extraction. For each method, we traverse its AST and extract the JavaDoc Comment part. We ignore methods without JavaDoc comments. 
	Then, we select the first sentence of the comment as the annotation.
	We exclude irregular annotations such as those starting with ``TODO: Auto-generated method stub'', ``NOTE:'', and ``test''. We also filter out non-words and words within brackets in the annotations.  
	
	Finally, we obtain a database consisting of 7,519,907 \textsf{$\langle$API sequence, annotation$\rangle$} pairs. 
	
	\subsection{Training Encoder-Decoder Language \\ Model}
	
	As described in Section~\ref{s:tech}, we adapt the attention-based RNN Encoder-Decoder model for API learning. The RNN has various implementations, we use GRU~\cite{cho2014phrase} which is a state-of-the-art RNN and performs well in many tasks~\cite{bahdanau2014attention,cho2014phrase}. We construct the model as follows: we use two RNNs for the encoder - a forward RNN that directly encodes the source sentence and a backward RNN that encodes the reversed source sentence. Their output context vectors are concatenated to the decoder, which is also an RNN. All RNNs have 1000 hidden units. We set the dimension of word embedding to 120. We discuss the details of parameter tuning in Section~\ref{ss:eval:para}. 
	
	All models are trained using the minibatch Adadelta~\cite{zeiler2012adadelta}, which automatically adjusts the learning rate.	
	 We set the batch size (i.e., number of instances per batch) as 200. For training the neural networks, we limit the source and target vocabulary to the top 10,000 words that are most frequently used in API sequences and annotations.
	
	For implementation, we use GroundHog~\cite{bahdanau2014attention,cho2014phrase}, an open-source deep learning framework. We train our models in a server with one Nvidia K20 GPU. The training lasts $\sim$240 hours with 1 million iterations. 
			
	\subsection{Translation}
		\begin{figure} [!tb]
			\setlength{\belowcaptionskip}{0pt}
			\centering 
			\includegraphics[width=3.3in]{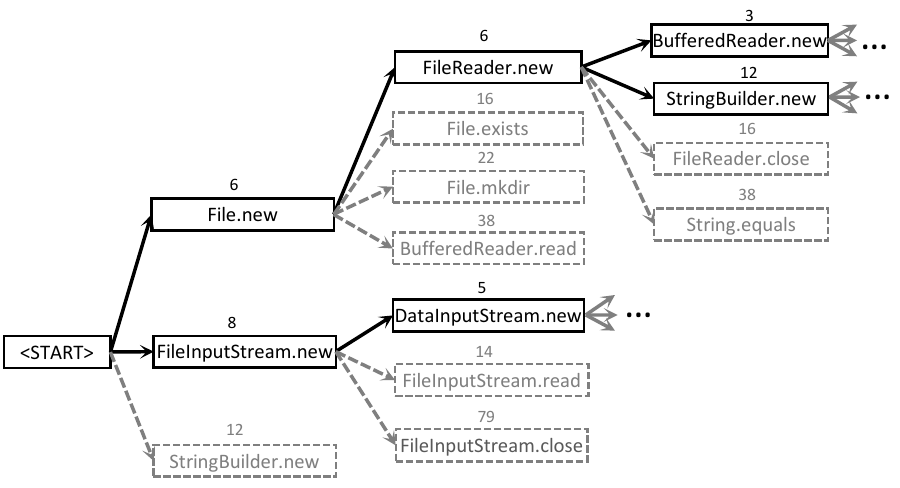} 
			\vspace{-20pt}
			\caption{\small An illustration of beam search (beam width=2)}
			\label{fig:approach:beamsearch}
			\vspace{-1.5\baselineskip}
		\end{figure}
	So far we have discussed the training of a neural language model, which outputs the most likely API sequence given a natural language query. However, an API could have multiple usages.
	To obtain a ranked list of possible API sequences for user selection, we need to generate more API sequences according to their probability at each step.
	
	\textsc{DeepAPI} uses Beam Search~\cite{beamsearch}, a heuristic search strategy, to find API sequences that have the least cost value(computed using Equation~\ref{eq:newcost}) given by the language model. 
	Beam search searches APIs produced at each step one by one. At each time step, it selects $n$ APIs from all branches with the least cost values, where $n$ is the beam-width. It then prunes off the remaining branches and continues selecting the possible APIs that follow on until it meets the end-of-sequence symbol. 
	Figure~\ref{fig:approach:beamsearch} shows an example of a beam search (beam-width=2) for generating an API sequence for the query ``read text file''. First, `START' is selected as the first API in the generated sequence. Then, it estimates the probabilities of all possible APIs that follow on according to the language model. It computes their cost values according to Equation~\ref{eq:newcost}, and selects \emph{File.new} and \emph{FileInputStream.new} which have the least cost values of 6 and 8, respectively. Then, it ignores branches of other APIs and continue estimating possible APIs after \emph{File.new} and \emph{FileInputStream.new}. Once it selects an end-of-sequence symbol as the next API, it stops that branch and the branch is selected as a generated sequence.   
	
	Finally, \textsc{DeepAPI} produces $n$ API sequences for each query where $n$ is the beam-width. We rank the generated API sequences according to their average cost values during the beam search procedure.


	\section{Evaluation}\label{s:eval}
	We evaluate the effectiveness of \textsc{DeepAPI} by measuring its accuracy on API sequence generation.
	Specifically, our evaluation addresses the following research questions:
	\begin{itemize}\setlength{\itemsep}{-\itemsep}
		\item \bf{RQ1: How accurate is \textsc{DeepAPI} for generating API usage sequences?}
		\item \bf{RQ2: How accurate is \textsc{DeepAPI} under different parameter settings?}
		\item \bf{RQ3: Do the enhanced RNN Encoder-Decoder models improve the accuracy of DeepAPI?}
	\end{itemize}

	\subsection{Accuracy Measure} 
	
	\subsubsection{Intrinsic Measure - BLEU}
	We use the BLEU score~\cite{bleu} to measure the accuracy of generated API sequences. The BLEU score measures how close a candidate sequence is to a reference sequence (usually a human written sequence). It is a widely used accuracy measure for machine translation in the machine learning and natural language processing literature~\cite{bahdanau2014attention,cho2014phrase,sutskever2014seq2seq}. In our API learning context, we regard a generated API sequence given a query as a candidate, and a human-written API sequence (extracted from code) for the same query as a reference. We use BLEU to measure how close the generated API sequence is to a human-written API sequence.
	
	Generally, BLEU measures the hits of n-grams of a candidate sequence to the reference.
	It is computed as:	
	\vspace{-0.2\baselineskip}
	\begin{equation}
		\small
		\vspace{-0.2\baselineskip}
		BLEU = BP\cdot \textrm{exp}(\sum_{n=1}^Nw_n\textrm{log}\,p_n)
	\end{equation}
	where each $p_n$ is the precision of n-grams, that is, the ratio of length $n$ subsequences in the candidate that are also in the reference:
	\begin{equation}
		\small
		p_n=\frac{\textrm{\# n-grams appear in the reference+1}}{\textrm{\# n-grams of candidate+1}} \quad\textrm{for}\; n=1,...,N
	\end{equation}
	where $N$ is the maximum number of grams we consider. We set $N$ to 4, which is a common practice in the Machine Learning literature~\cite{sutskever2014seq2seq}.
	Each $w_n$ is the weight of each $p_n$. A common practice is to set $w_n=\frac{1}{N}$.
	$BP$ is a brevity penalty which penalties overly short candidates (that may have a higher n-gram precision).
	\begin{equation} 
		\small
		\vspace{-0.2\baselineskip}
	   BP =
	\begin{cases}
	1   &  \textrm{if} \; c > r \\
	e^{(1-r/c)}   &  \textrm{if} \; c \leq r
	\end{cases}                
	\end{equation}
	where $r$ is the length of the reference sequence, and $c$ is the length of the candidate sequence.

	We now give an example of BLEU calculation. For a candidate API sequence \{a-c-d-b\} and a reference API sequence \{a-b-c-d-e\}, their 1-grams are \{a,b,c,d\} and \{a,b,c,d,e\}. All four 1-grams of the candidate are hit in the reference. Then, $p_1=\frac{4+1}{4+1}=1$. Their 2-grams are \{ac,cd,db\} and \{ab,bc,cd,de\}, respectively. Then, $p_2=\frac{1+1}{3+1}=\frac{1}{2}$ as only $cd$ is matched. $p_3=\frac{0+1}{2+1}=\frac{1}{3}$ and $p_4=\frac{0+1}{1+1}=\frac{1}{2}$ as no 3-gram nor 4-gram is matched. As their lengths are 4 and 5 respectively, $BP=e^{(1-5/4)}=0.78$. The final BLEU is $0.78\times exp(\frac{1}{4}\times log1+\frac{1}{4}\times log\frac{1}{2}+\frac{1}{4}\times log\frac{1}{3}+\frac{1}{4}\times log\frac{1}{2})=41.91\%$
	
	BLEU is usually expressed as a percentage value between 0 and 100. The higher the BLEU, the closer the candidate sequence is to the reference. If the candidate sequence is completely equal to the reference, the BLEU becomes 100\%.
	
	\subsubsection{Extrinsic Measures - FRank and Relevancy Ratio}
	We also use two measures for human evaluation. They are FRank and relevancy ratio~\cite{swim}.
	FRank is the rank of the first relevant result in the result list~\cite{swim}. It is important as most users scan the results from top to bottom. 
	
	The relevancy ratio is defined as the precision of relevant results in a number of results~\cite{swim}. 
	\vspace{-0.2\baselineskip}
	\begin{equation}
	\small
	\textrm{relevancy ratio}=\frac{\# \textrm{relevant results}}{\# \textrm{all selected results}}
	\end{equation}
	\vspace{-0.8\baselineskip}
	
	The value of both measures ranges from 0 to 100. The higher the better. 

	
	\subsection{Comparison Methods}
	We compare the accuracy of our approach with that of two state-of-the-art API learning approaches, namely Code Search with Pattern Mining ~\cite{sourcerer,upminer}
	and SWIM~\cite{swim}.
	\subsubsection{Code Search with Pattern Mining} 
	To obtain relevant API sequences for a given a query, one can perform code search over the code corpus using information retrieval techniques~\cite{exoa,sourcerer, codehow,portfolio}, and then utilize an API usage pattern miner~\cite{fowkes2015parameter,upminer,mapo2006} to identify an appropriate API sequences in the returned code snippets.
		
	We compare DeepAPI with this approach. We use Lucene \cite{lucene} to perform a code search for a given natural language query and UP-Miner~\cite{upminer} to perform API usage pattern mining. Lucene is an open-source information retrieval engine, which has been integrated into many code search engines~\cite{sourcerer,swim}. Much the same as these code search engines do, we treat source code as plain text documents and use Lucene to build source code index and perform text retrieval. 
	UP-Miner~\cite{upminer} is a pattern mining tool, which produces API sequence patterns from code snippets. It first clusters API sequences extracted from code snippets, and then identifies frequent patterns from the clustered sequences. Finally, it clusters the frequent patterns to reduce redundancy. We use UP-Miner to mine API usage sequences from the code snippets returned by the Lucene-based code search. 
	
	In this experiment, we use the same code corpus as used for evaluating \textsc{DeepAPI}, and compare the BLEU scores with those of \textsc{DeepAPI}.

	\subsubsection{SWIM} 
	SWIM~\cite{swim} is a recently proposed code synthesis tool, which also supports API sequence search based on a natural language query. Given a query, it expands the query keywords to a list of relevant APIs using a statistical word alignment model~\cite{smt}. With the list of possible APIs, SWIM searches related API sequences using Lucene~\cite{lucene}. Finally, it synthesizes code snippets based on the API sequences.
	As code synthesis is beyond our scope, we only compare \textsc{DeepAPI} with the API learning component of SWIM, that is, from a natural language query to an API sequence. In their experiments, SWIM uses Bing clickthrough data to build the model. In our experiment, for fair comparison, we evaluate SWIM using the same dataset as we did for evaluating \textsc{DeepAPI}. 
	That is, we train the word alignment model and build API index on the training set, and evaluate the search results on the test set.	
	
	\subsection{Accuracy (RQ1)}\label{ss:eval:rq1}
	
	\subsubsection{Intrinsic Evaluation} 
	
	\textbf{Evaluation Setup:}
	We first evaluate the accuracy of generated API sequences using the BLEU score.
	As described in Section~\ref{ss:approach:data}, we collect a database comprising 7,519,907 \textsf{$\langle$API sequence, annotation$\rangle$} pairs. We split them into a test set and a training set. The test set comprises 10,000 pairs while the training set consists of the remaining instances.
	We train all models using the training set and compute the BLEU scores in the test set. 	
	We calculate the highest BLEU score for each test instance in the top $n$ results. 
	
	\textbf{Results:}
	Table~\ref{tab:result:bleucomp:approach} shows the BLEU scores of \textsc{DeepAPI}, SWIM, and Code Search (Lucene+UP-Miner). Each column shows the average BLEU score for a method. As the results indicate, \textsc{DeepAPI} produces API sequences with higher accuracy. When only the top 1 result is examined, the BLEU score achieved by \textsc{DeepAPI} is 54.42, which is greater than that of SWIM (BLEU=19.90) and Code Search (BLEU=11.97). The improvement over SWIM is 173\% and the improvement over Code Search is 355\%.
	Similar results are obtained when the top 5 and 10 results are examined. 
	The evaluation results confirm the effectiveness of the deep learning method used by \textsc{DeepAPI}.
	
	\begin{table}[tb]
		\centering
		\scriptsize
		\caption{\small BLEU scores of \textsc{DeepAPI} and related techniques (\%)}
		\begin{threeparttable}
			\begin{tabular}{p{2.3cm}|p{0.6cm}|p{0.6cm}|p{0.7cm}} 
				\toprule[1pt]
				\rowcolor{gray!15}
				Tool				&    Top1	&	Top5   	& Top10 	 \\ \hline
				Lucene+UP-Miner		&  	 11.97		& 24.08		& 	29.64	\\ 
				SWIM				& 	19.90  	&  25.98   	& 	28.85		\\ \hline
				\textsc{DeepAPI}	&  	54.42	&  64.89 	& 	67.83		\\
				\bottomrule[1pt] 
			\end{tabular}
		\end{threeparttable}
		\label{tab:result:bleucomp:approach}
		\vspace{-1.5\baselineskip}
	\end{table}


	
	\subsubsection{Extrinsic Evaluation}
	To further evaluate the relevancy of the results returned by \textsc{DeepAPI}, we selected 17 queries used in~\cite{swim}. These queries have corresponding Java APIs and are commonly occurring queries in the Bing search log~\cite{swim}. To demonstrate the advantages of \textsc{DeepAPI}, we also designed 13 longer queries and queries with semantically similar words. In total, 30 queries are used. These queries do not appear in the training set. Table~\ref{tab:eval:extrinsic} lists the queries. 
	\begin{table*}[tb]
		\scriptsize
		\centering
		\caption{\small Queries for Human Evaluation (FR: FRank, RR5: top 5 relevancy ratio, RR10: top 10 relevancy ratio)}
		\begin{tabular}{@{}p{3.2cm}|@{}c@{}|@{}c@{}|@{}c@{}|@{}c@{}|@{}c@{}|@{}c@{}|p{9.7cm}@{}} 
			\hline
			\multirow{2}{*}{\bf query (How to...)} & \multicolumn{3}{c|}{ SWIM} & \multicolumn{3}{c|}{ DeepAPI} & \multirow{2}{*}{\bf Generated API sequence by DeepAPI}\\ \cline{2-7}
			&  \;\;FR\;\; & \;RR5\; & RR10 & \;\;FR\;\; & \;RR5\; & RR10 &   \\ \hline
			convert int to string 	&8&0&10&2	& 40 & 90&	Integer.toString  \\ 
			convert string to int 	&1	& 80 & 80 &1&100&100& Integer.parseInt String.toCharArray Character.digit\\
			append strings &3 &60 &80 &1&100&100 & StringBuilder.append StringBuilder.toString\\
			get current time &	1& 80 & 80&10&10&10& System.currentTimeMillis Timestamp.new  \\
			parse datetime from string &9 &0 &10 & 1&100 &80 & SimpleDateFormat.new SimpleDateFormat.parse\\
			test file exists & - &0 &0& 1 & 100&100& File.new File.exists \\
			open a url & 1 & 100 & 100 & 1&100&100& URL.new URL.openConnection\\
			open file dialog &-&0&0&1&100&80&JFileChooser.new JFileChooser.showOpenDialog JFileChooser.getSelectedFile \\
			get files in folder &2 &40&20& 3 & 40 & 50 & File.new File.list File.new File.isDirectory \\
			match regular expressions &1 & 100& 100& 1& 80& 90& Pattern.compile Pattern.matcher Matcher.group \\
			generate md5 hash code &1& 60&40 &1&100&100& MessageDigest.getInstance MessageDigest.update MessageDigest.digest\\
			generate random number &7&0 & 10 &1&100&70& Random.new Random.nextInt  \\
			round a decimal value &- &0&0  &1&100&100&Math.floor  Math.pow  Math.round \\
			execute sql statement & 2 &80&80  &1&80&60&Connection.prepareStatement PreparedStatement.execute PreparedStatement.close \\
			connect to database&7&0 &20&1&100&90 & Properties.getProperty Class.forName DriverManager.getConnection \\
			create file &10 & 0&10&3&40&20&File.exists File.createNewFile \\
			copy file &1&100&100&2&20&10&FileInputStream.new FileOutputStream.new FileInputStream.read FileOutputStrem.write FileInputStream.close FileOutputStream.close\\
			copy a file and save it to -your destination path &1 &20 & 50& 1&100 &100 & FileInputStream.new FileOutputStream.new FileInputStream.getChannel FileOutputStream.getChannel FileChannel.size FileChannel.transferTo FileInputStream.close FileOutputStream.close FileChannel.close FileChannel.close\\
			delete files and folders in a -directory &1&100&90&1&100&100 & File.isDirectory File.list File.new File.delete\\
			reverse a string &3&20&10&2&60&70& StringBuffer.new StringBuffer.reverse \\
			create socket&-&0&0&1&60&80&ServerSocket.new ServerSocket.bind\\
			rename a file &- &0 & 0& 1&100 & 100&File.renameTo File.delete\\
			download file from url & 2& 60& 80&1 &100 &80 & URL.new URL.openConnection URLConnection.getInputStream BufferedInputStream.new\\
			serialize an object &1 &100 &100 & 3&60 &70 & ObjectOutputStream.new ObjectOutputStream.writeObject ObjectOutputStream.close\\
			read binary file & 4& 40&70 &1 &100 &80 & DataInputStream.new DataInputStream.readInt DataInputStream.close\\
			save an image to a file &1&20&10 &1&80&80& File.new ImageIO.write \\
			write an image to a file&1&20&10&1&100&90& File.new ImageIO.write \\
			parse xml &1 &100 &100&1&80&60& InputSource.new DocumentBuilder.parse\\
			play audio &1&100&100&1&60&80&SourceDataLine.open SourceDataLine.start\\ 
			play the audio clip at the -specified absolute URL&1&40&50&1&100&90&Applet.getAudioClip AudioClip.play\\
			\hline
			\bf average &>4.0 &44&47&1.6&80&78&\\ \hline
		\end{tabular}
		\label{tab:eval:extrinsic}
		\vspace{-1.5\baselineskip}
	\end{table*}
	
	For each query, the top 10 returned results by \textsc{DeepAPI} and SWIM are manually examined. To reduce labeling bias, two developers separately label the relevancy of each resulting sequence and combine their labels. For inconsistent labels, they discuss and relabel them until a settlement is reached.  
	The FRank and the relevancy ratios for the top 5 and top 10 returned results are then computed. To test the statistical significance, we apply the Wilcoxon signed-rank test ($p$$<$0.05) for all results of both approaches. A resulting p-value less than 0.05 indicates that the differences between \textsc{DeepAPI} and SWIM are statistically significant.
	 
	Table~\ref{tab:eval:extrinsic} shows the accuracy of both \textsc{DeepAPI} and SWIM. The symbol `-' means no relevant result has been returned within the top 10 results. 
	The results show that \textsc{DeepAPI} is able to produce mostly relevant results. It achieves an average FRank of 1.6, an average top 5 accuracy of 80\%, and an average top 10 accuracy of 78\%. Furthermore, \textsc{DeepAPI} produces more relevant API sequences than SWIM, whose average top 5 and top 10 accuracy is 44\% and 47\%, respectively. For some queries, SWIM failed to obtain relevant results in the top 10 returned results. We conservatively treat the FRank as 11 for these unsuccessful queries. Then, the FRank achieved by SWIM is greater than 4.0, which is much higher than what \textsc{DeepAPI} achieved (1.60). The p-values for the three comparisons are 0.01, 0.02 and 0.01, respectively, indicating statistical significance of the improvement of \textsc{DeepAPI} over SWIM. In summary, the evaluation results confirm the effectiveness of \textsc{DeepAPI}. 
	
	Table~\ref{tab:eval:extrinsic} also shows examples of generated sequences by \textsc{DeepAPI}. We can see that
	\textsc{DeepAPI} is able to distinguish word sequences. For example, \textsc{DeepAPI} successfully distinguishes the query \emph{convert int to string} from \emph{convert string to int}.
	Another successful example is the query expansion. For example, the query \emph{save an image to a file} and \emph{write an image to a file} return similar results. 
	\textsc{DeepAPI} also performs well in longer queries such as \emph{copy a file and save it to your destination path} and \emph{play the audio clip at the specified absolute URL}.
	Such queries comprise many keywords, and \textsc{DeepAPI} can successfully recognize the semantics.
	
	We also manually check the results returned by SWIM. We find SWIM may return partially matched sequences. 
	For example, for the query \emph{generate md5 hash code}, SWIM returns many results containing only \emph{Object.hashCode}, which simply returns a hash code. SWIM also returns project specific results without fully understanding the query. For example, for the query ``test file exists'', SWIM returns ``\textit{File.new, File.exists, File.getName, File.new, File.delete, FileInputStream.new, FileInputStream.read, ...''}, which is not only related to file existence test, but also to other project-specific tasks. Such project specific results can also be seen for the query \emph{create file}. 
	Compared with DeepAPI, SWIM performs worse in long queries. For example, SWIM performs worse in the query \emph{copy a file and save it to your destination path} than in the query \emph{copy file}. This is because long queries often have multiple objectives, which cannot be understood by SWIM. 
	
	Still, \textsc{DeepAPI} could return inaccurate or partial results. For example, for the query \emph{parse xml}, it returns related APIs \emph{InputSource.new, DocumentBuilder.parse}. But it misses the APIs about how \emph{DocumentBuilder} is created (\emph{DocumentBuilderFactory.newDocumentBuilder}). The reason could be that an API sequence may be called in an inter-procedural manner. When preparing the training set, we only consider API sequences within one method. The API \emph{DocumentBuilderFactory.newDocumentBuilder} could be called in another method and is passed as a parameter. This causes incomplete sequences in the training set. In the future, we will perform more accurate program analysis and create a better training set.

	\vspace{-0.3\baselineskip}
	\subsection{Accuracy Under Different Parameter Settings (RQ2)}\label{ss:eval:para}
		We also qualitatively compare the accuracy of \textsc{DeepAPI} in different parameter settings. 
		We analyze two parameters, that is, the dimension of word embedding and the number of hidden units.
		We vary the values of these two parameters and evaluate their impact on the BLEU scores.
		
\begin{figure}[tb]
			\setlength{\abovecaptionskip}{1pt}
			\setlength{\belowcaptionskip}{0pt}
			\centering
			\begingroup
				\captionsetup[subfigure]{width=3in}
				\subfloat[\scriptsize Performance of different dimensions of word embedding]{
					\includegraphics[width=2in]{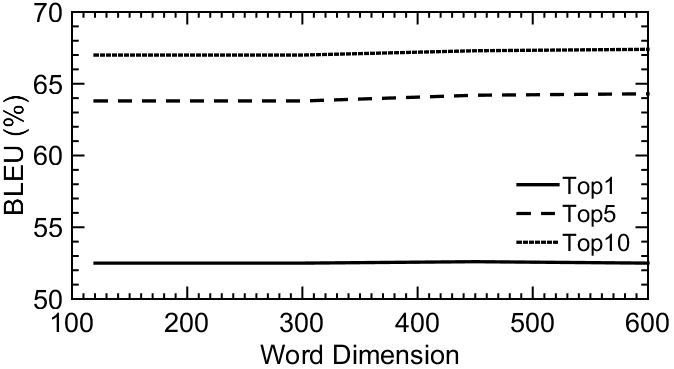}
					\label{fig:result:paracomp:emb}} \\
				\subfloat[\scriptsize Performance of different numbers of hidden units]{
					\includegraphics[width=2in]{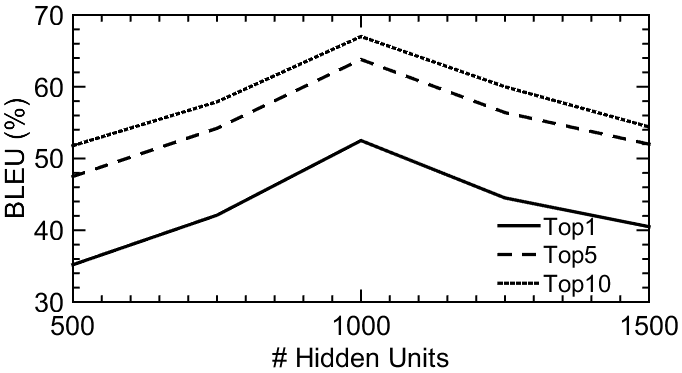}
					\label{fig:result:paracomp:hid}}
			\endgroup
			\vspace{-8pt}
			\caption{\small BLEU scores of different parameter settings}
			\label{fig:result:paracomp}
			\vspace{-16pt}
		\end{figure}

		Figure~\ref{fig:result:paracomp} shows the influence of different parameter settings on the test set. The dimension of word embedding makes little difference to the accuracy.
		The accuracy of \textsc{DeepAPI} greately depends on the number of hidden units in the hidden layer. The optimum number of hidden units is around 1000.   
		
%
%

	\subsection{Performance of the Enhanced RNN Encoder-Decoder Models (RQ3)}
	
			In Section~\ref{s:tech}, we describe two enhancements to the original RNN Encoder-Decoder model, for the task of API learning: an attention-based RNN Encoder-Decoder proposed by \cite{bahdanau2014attention} (Section~\ref{ss:tech:attension}) and an enhanced RNN Encoder-Decoder with a new cost function (Section~\ref{ss:tech:enhance}) proposed by us. We now evaluate if the enhanced models improve the accuracy of \textsc{DeepAPI} when constructed using the original RNN Encoder-Decoder model.
			
			Table~\ref{tab:result:bleucomp:scheme} shows the BLEU scores of the three models. The attention-based RNN Encoder-Decoder outperforms the basic RNN Encoder-Decoder model on API learning. The relative improvement in the top 1, 5, and 10 results (in terms of BLEU score) is 8\%, 5\% and 4\%, respectively. This result confirms the effectiveness of the attention-based RNN Encoder-Decoder used in our approach.
	
			Table~\ref{tab:result:bleucomp:scheme} also shows that the enhanced model with the new cost function leads to better results as compared to the attention-based RNN Encoder-Decoder model. The improvement in the top 1, 5, and 10 results (in terms of BLEU score) is 4\%, 2\% and 1\%, respectively.
			Figure~\ref{fig:result:enhance} shows that the performance of the enhanced model are slightly different under different parameter settings, with an optimum $\lambda$ of around 0.035. The results confirm the usefulness of the proposed cost function for enhancing the RNN Encoder-Decoder model.
		\begin{table}[tb]
			\setlength{\abovecaptionskip}{0pt}
			\setlength{\belowcaptionskip}{0pt}
			\centering
			\scriptsize
			\caption{\small BLEU scores of different RNN Encoder-Decoder Models (\%)}
			\begin{threeparttable}
				\begin{tabular}{p{4.6cm}|p{0.8cm}|p{0.8cm}|p{0.8cm}} 
					\toprule[1pt]
					\rowcolor{gray!15}
					Encoder-Decoder Model	&    Top1	&	Top5   	& Top10 	 \\ \hline
					RNN				&  	48.83	& 60.98		& 	64.27				\\ 
					RNN+Attention &  	52.49	& 63.81  	& 	66.97				\\
					RNN+Attention+New Cost Function& 	54.42 	&  	64.89	& 67.83	\\ 
					\bottomrule[1pt] 
				\end{tabular}
			\end{threeparttable}
			\label{tab:result:bleucomp:scheme}
			\vspace{-1\baselineskip}
		\end{table}
		
		\begin{figure}[tb]
			\setlength{\abovecaptionskip}{0pt}
			\setlength{\belowcaptionskip}{0pt}
			\centering
			\includegraphics[width=2in]{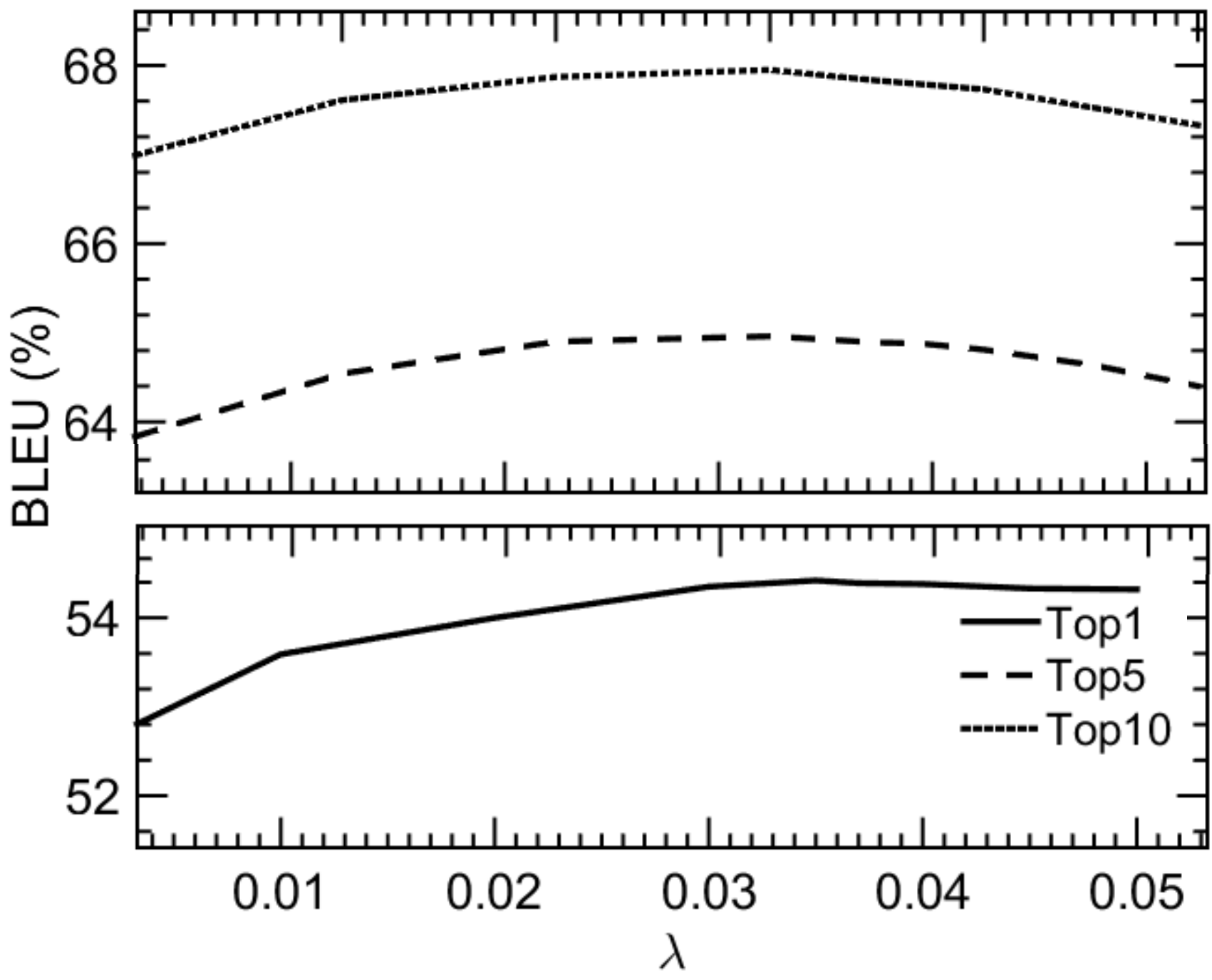}
			\vspace{-8pt}
			\caption{\small Performance of the Enhanced RNN Encoder-Decoder Model under Different Settings of $\lambda$}
			\label{fig:result:enhance}
			\vspace{-1\baselineskip}
		\end{figure}

	\section{Discussion}\label{s:discuss}
	\subsection{Why does DeepAPI work?}
	A major challenge for API learning is the semantic gap between code and natural language descriptions. Existing information retrieval based approaches usually have a bag-of-words assumption and lack a deep understanding of the high-level semantics of natural language and code. We have identified three advantages of \textsc{DeepAPI} that address this problem.
	\begin{figure*} [t]
		\setlength{\abovecaptionskip}{0pt}
		\centering 
		\includegraphics[width=7in]{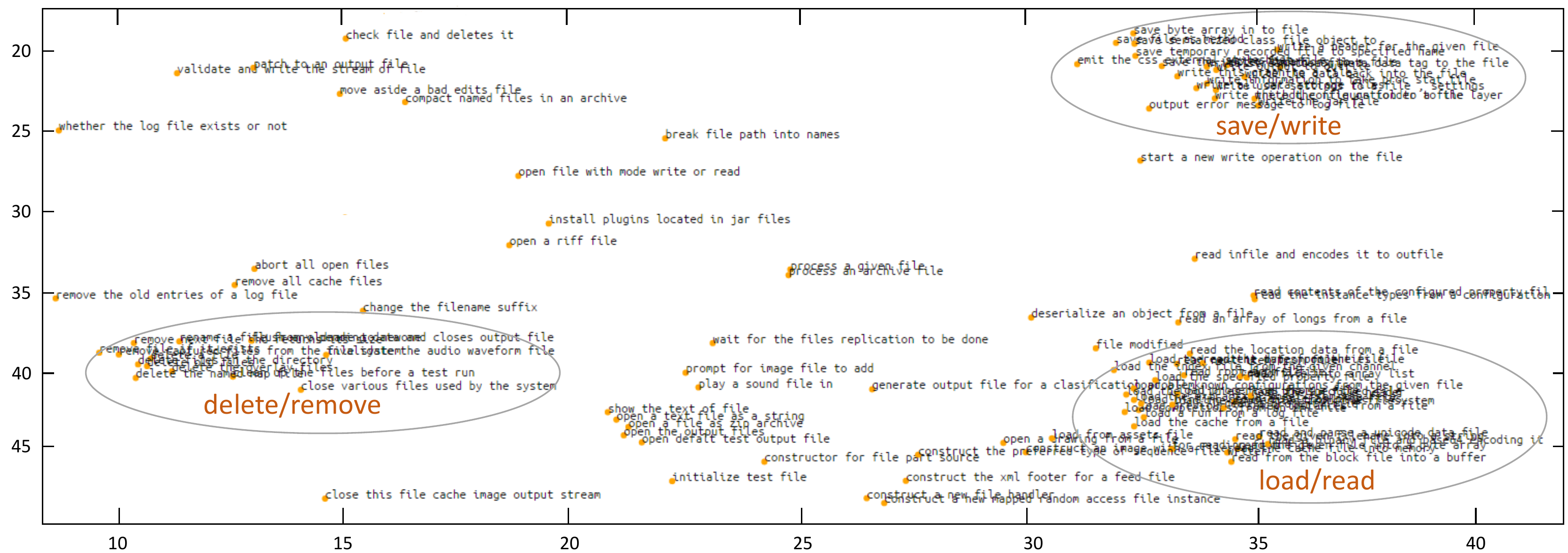} 
		\vspace{-20pt}
		\caption{\small A 2D projection of embeddings of queries using t-SNE\cite{tsne}}
		\label{fig:discuss:queryembed}
		\vspace{-1\baselineskip}
	\end{figure*}	
		\\ \textbf{Word embedding and query expansion}
		A significant difference between \textsc{DeepAPI} and bag-of-words methods is, \textsc{DeepAPI} embeds words into a continuous semantic space where the semantically similar words are placed close to each other. When reading words in a query, the model maps them to semantic vectors. Words with similar semantics have similar vector representations and have a similar impact on the hidden states of the RNN encoder. Therefore, queries with semantically similar words can lead to similar results. 
		Figure~\ref{fig:discuss:queryembed} shows a 2-D projection of the encoded vectors of queries. These queries are selected from the 10,000 annotations in the test set. For ease of demonstration, we select queries with a keyword ``file'' and exclude those longer than eight words. As shown in the graph, \textsc{DeepAPI} can successfully embed similar queries into a nearby place. 
		There are three clear clusters of queries, corresponding to ``read/load files'',``write/save files'', and ``remove/delete files''. Queries with semantically related words are close to each other. For example, queries starting with \emph{save}, \emph{write}, and \emph{output} are in the same ``cluster'' though they contain different words.  		
		\\ \textbf{Learning sequence instead of bag-of-words}
		The hidden layer of the encoder has the memory capacity. 
		It considers not only the individual words, but also their relative positions. Even for the same word set, different sequences will be encoded to different vectors, resulting in different API sequences. 
		In that sense, \textsc{DeepAPI} learns not only the words, but also phrases. While traditional models simply consider individual words or word-level alignments. 
		A typical example is that, queries with different word sequences such as \emph{convert int to string} and \emph{convert string to int} can be distinguished well by \textsc{DeepAPI}. 
		\\\textbf{Generating common patterns instead of searching specific samples}
		Another advantage of our approach is that, it can learn common patterns of API sequences. The decoder itself is a language model and remembers the likelihoods of different sequences. Those common sequences will have high probabilities according to the model. Therefore, it tends to generate common API sequences rather than project-specific ones. On the other hand, the information retrieval based approaches simply consider searching individual instances and could return project-specific API sequences. 
	
	Though several techniques such as query expansion~\cite{howard2013automatically,shepherd2007using,yang2012inferring} and frequent pattern mining~\cite{mapo2006} can partially solve some of the above problems, their effectiveness remains to be improved. For example, it has been observed that expanding a code	search query with inappropriate English synonyms can return even worse results as compared to the original query~\cite{sridhara2008identifying}. Furthermore, few techniques can exhibit all the above advantages.

	\subsection{Threats to Validity}\label{ss:threats}
	We have identified the following threats to validity:\\
	\textbf{All APIs studied are Java APIs} 
	All APIs and related projects investigated in this paper are JDK APIs. Hence, they might not be representative of APIs for other libraries and programming languages. In the future, we will extend the model to other libraries and programming languages. 
	\\\textbf{Quality of annotations} We collected annotations of API sequences from the first sentence of  documentation comments. Other sentences in the comments may also be informative. In addition, the first sentences may have noise. In the future, we will investigate a better NLP technique to extract annotations for code.
	\\
	\textbf{Training dataset} In the original SWIM paper~\cite{swim}, the clickthrough data from Bing.com is used for evaluation. Such data is not easy accessible for most researchers. 
	For fair and easy comparison, we evaluate SWIM on the dataset collected from GitHub and Java documentations (the same for evaluating \textsc{DeepAPI}). We train the models using annotations of API sequences collected from the documentation comments. In the future, we will evaluate both SWIM and \textsc{DeepAPI} on a variety of datasets including the Bing clickthrough data.  
	In the future, we will perform more accurate program analysis and create a better training set.  

	\section{Related Work}\label{s:related}
	
	\vspace{-6pt}
	\subsection{Code Search}
	There is a large amount of work on code search~\cite{codesearch,chan2012searching,holmes2009end,keivanloo2014spotting,codehow,portfolio}. For example, McMillan et al.~\cite{portfolio} proposed a code search tool called Portfolio that retrieves and visualizes relevant functions and their usages. 
	Chan and Cheng~\cite{chan2012searching} designed an approach to help users find usages of APIs given only simple text phrases. 
	Lv et al.~\cite{codehow} proposed CodeHow, a code search tool that incorporates an extended Boolean model and API matching. They first find relevant APIs to a query by matching the query to API documentation. Then, they improve code search performance by considering the APIs which are relevant to the query in code retrieval. As described in Section~\ref{s:discuss}, \textsc{DeepAPI} differs from code search techniques in that it does not rely on information retrieval techniques and can understand word sequences and query semantics.  
			
	\subsection{Mining API Usage Patterns}
	Instead of generating API sequences from natural language queries, there is a number of techniques focusing on mining API usage patterns~\cite{fowkes2015parameter,moritz2013export,upminer,mapo2006}. API usage patterns are frequent API method call sequences. Xie et al.~\cite{mapo2006} proposed MAPO, which is one of the first works on mining API patterns from code corpus. MAPO represents source code as call sequences and clusters them according to similarity heuristics such as method names. It finally generates patterns by mining and ranking frequent sequences in each cluster. UP-Miner~\cite{upminer} is an improvement of MAPO, which removes the redundancy among patterns by two rounds of clustering of the method call sequences. 
	By applying API usage pattern mining on large-scale code search results, these techniques can also return API usage sequences in response to user's natural language queries. 
	
	While the above techniques are useful for understanding the usage of an API, they are insufficient for answering the question of \textit{which APIs to use}, which is the aim of \textsc{DeepAPI}. Furthermore, different from a frequent pattern mining approach, \textsc{DeepAPI} constructs a neural language model to learn usage patterns.
	
	\subsection{From Natural Language to Code}
	A number of related techniques have been proposed to generate code snippets from natural language queries. For example, Raghothaman et al.~\cite{swim} proposed SWIM, a code synthesis technique that translates user queries into the APIs of interest using Bing search logs and then synthesizes idiomatic code describing the use of these APIs. SWIM has a component that produces API sequences given user's natural language query. Our approach and SWIM differ in many aspects. First, SWIM generates bags of APIs using statistical word alignment~\cite{smt}. The word alignment model does not consider word embeddings and word sequences of natural language queries, and has limitations in query understanding. Second, to produce API sequences, SWIM searches API sequences from the code repository using a bag of APIs. It does not consider the relative position of different APIs. 
	Fowkes and Sutton~\cite{allamanis2015bimodal} build probabilistic models that jointly model short natural language utterances and source code snippets. The main differences between our approach and theirs are two-fold. First, they use a bag-of-words model to represent natural language sentences which will not recognize word sequences. Second, they use a traditional probabilistic model which is unable to recognize semantically related words. 
	
	
	\subsection{Deep Learning for Source Code}
	Recently, some researchers have explored the possibility of applying deep learning techniques to source code~\cite{allamanis2016convolutional,mou2014vec,mou2016cnn,rnncodecompletion}. 
	A typical application that leverages deep learning is to extract source code features~\cite{mou2014vec,wang2016automatically}. For example, Mou et al.~\cite{mou2014vec} proposed to learn vector representations of source code for deep learning tasks. Mou et al.~\cite{mou2016cnn} also proposed convolutional neural networks over tree structures for programming language processing. 
	Deep learning has also been applied to code generation~\cite{ling2016latent,mou2015codegen}. For example, Mou et al.~\cite{mou2015codegen} proposed to generate code from natural language user intentions using an RNN Encoder-Decoder model. Their results show the feasibility of applying deep learning techniques to code generation from a highly homogeneous dataset (simple programming assignments). 
	Deep Learning has also been applied to code completion~\cite{rnncodecompletion,white2015toward}. For example, White et al.~\cite{white2015toward} applied the RNN language model to source code files and showed its effectiveness in predicting software tokens.
	Raychev et al.~\cite{rnncodecompletion} proposed to apply the RNN language model to complete partial programs with holes. 
	In our work, we explore the application of deep learning techniques to API learning. 

	\section{Conclusion}\label{s:conclusion}
	In this paper, we apply a deep learning approach, RNN Encoder-Decoder, for generating API usage sequences for a given API-related natural language query. 
	Our empirical study has shown that the proposed approach is effective in API sequence generation.
	Although deep learning has shown promise in other areas,
	we are the first to observe its effectiveness in API learning. 
	
	The RNN Encoder-Decoder based neural language model described in this paper may benefit other software engineering problems such as code search and bug localization. In the future, we will explore the applications of this model to these problems. We will also investigate the synthesis of sample code from the generated API sequences.
	
	An online demo of \textsc{DeepAPI} can be found on our website at: \url{https://guxd.github.io/deepapi/}.


	\balance
	\bibliographystyle{abbrv}
	\bibliography{references} 
	
\end{document}